\documentclass[11pt,twoside]{article}  
\usepackage{adassconf}

\begin{document}   

\paperID{P2.01}

\title{A Customizable Database Server:
 MCS, a Flexible Resource for Astronomical Projects}
\titlemark{A Customizable Database Server}

\author{Luciano Nicastro}
\affil{INAF--IASF,
    Via P. Gobetti 101, 40129 Bologna, Italy}
\author{Giorgio Calderone}
\affil{INAF--IASF,
    Via U. La Malfa 153, 90146 Palermo, Italy}

\contact{Luciano Nicastro}
\email{nicastro@iasfbo.inaf.it}

\paindex{Nicastro, L.}
\aindex{Calderone, G.}

\authormark{Nicastro \& Calderone}

\keywords{databases: customization, data: management}

\begin{abstract}
Nowadays medium-large size astronomical projects have to face the management
of a large amount of information and data. Dedicated data centres
manage the collection of raw and processed data and consequently make
them accessible, typically as static files, either via (s)ftp or http (web).
However the various steps of data acquisition, archiving, processing and
delivering are accomplished by different tools.
The data production information (logging) is not always collected into
relational databases tables causing long delays before people know about
their existence.
Luckily the use, in many cases, of standard file formats like FITS can help
to track the data origin and processing status. As Virtual Observatory
standards will become more common, things will quickly improve.
Here we present a software library we implemented in order to use a unified
model in astronomical data treatment. All the data are stored into the same
database becoming available in different
forms to different users with different privileges.
\end{abstract}

\section{Introduction}
Information services can be separated in two classes: those in which
the information produced is addressed to humans, and those in which
it is meant to be used by other software applications. In the
former case there is a quite standardized way to develop such an
information service, essentially based upon a web server, a database (DB)
server, a scripting language and HTML pages. In the latter case
instead there is no such standardization, and here's why
\textbf{MCS} (My Customizable Server) was implemented (Calderone \& Nicastro
2006).

MCS is a set of software tools aimed at easily implementing information
services, that is an application providing a service over the
network. At the core of an MCS-based system there is a TCP server listening
for user connections requests. When a user is connected, the server will
send him
all the requested information. However the transmitted data aren't in
free format (like in a web page), but are packed in a well defined
fashion (using the MCS protocol) so that on the other side a software
can understand what is being sent. The MCS high level classes will
hide all code implementations related to multi-threading, networking,
DB access, etc., and require no low-level knowledge of these
issues by the users. MCS, as its name suggests, can be customized
through the derivation of some classes.
So MCS and its protocol are for software applications what a web
server and HTTP are for the WWW: a simple way to access any sort of data.
In this comparison customizing the MCS server is like writing a web page.
The usage of MCS for the automatic management of the data
collected by the robotic telescope REM is described in Nicastro \& Calderone
(2006).

MCS is developed on the GNU/Linux platform and it is an ongoing project.
It is released under the GPL license and can be freely downloaded from the
\htmladdnormallink{web}{ross.iasfbo.inaf.it/mcs/}.
The site contains all news, updates, documentation and downloadable software
packages.

\section{The MCS Architecture and Components}
The main features of application servers built with MCS are:
 easy to configure,
 authentication and grant support,
 secure connections (through SSL),
 files transfer,
 DB access (MySQL),
 base commands set, support to create new customized commands,
 accessibility (as client) from other languages (C,
  Fortran, IDL, PHP, Python, etc.),
 XML, FITS and VOTAble management,
 HTM and HEALPix sky pixelization support,
 logging facility, etc.
\\
All these features are already available, without performing any
customization, and more will be added in the future.
So to implement a simple service with the above
features you'll only need to install MCS and configure it through a
simple configuration file. 
The only code needed for the ``simplest'' server is as follows:
\begin{verbatim}
#include <mcs.hh>                    //Mandatory includes for all
using namespace mcs;                 //  MCS-based applications
int main(int argc, char *argv[]) {   //Main program
  Env* env = mcsStart("simplest");   //Start the server...
  mcsWait(env);                  }   //...and wait for its end
\end{verbatim}
For more complex services you can customize the server behavior in
several ways:

\begin{itemize}
\item adding external programs, either real external applications or
  batch lists of MCS commands;
\item adding SQL programs, to be executed on the DB server;
\item adding customized commands, deriving the \verb|UserThread|
  class;
\item modifying the behavior of the server side thread, deriving the
  \verb|LocalThread| class;
\end{itemize}

In a typical application you should implement a DB with all
the tables needed to store the relevant data related
to a project, eventually preparing the required external programs.
A typical architecture of an MCS-based system is:
\\ \textbf{Database server}: the DB server is used to handle
clients authentication, to store all application specific data and
anything else necessary to the application itself. This server isn't
accessible directly from the clients, but it is visible only to the
application server. At the moment the only supported DB server is
MySQL.  
Other servers could become accessible in the future.
\\ \textbf{Application server}:
the application server is the core of the information system. It
implements the client/server model: a client opens a TCP socket
towards the host running MCS and sends a request, then the server
``computes'' an answer, eventually querying the DB and/or
executing some external programs, and sends it back to the client.
The behavior of the MCS server can be customized.
\\ \textbf{External programs}: external programs are
software applications, written in any language, which interact with the
application server via command line and the standard output. Support
to these programs was added to easily integrate in MCS already existing
applications.
\\ \textbf{Clients}: clients are programs
which access the MCS service over the network. Such programs can be
written in any language and run on any platform, provided that they
implement the MCS protocol. Interfaces that implement the MCS protocol
are provided by the MCS library for the following languages on the
Linux platform: C++, C, Fortran, IDL, PHP, Python.
Support for other languages, such as Java, Perl and Tcl/Tk, and the
Windows platform will be available soon.

MCS-based systems are ``open'', in the sense that
existing DB and software tools can be easily integrated.
Moreover MCS doesn't make any assumption
about type and quantity of data you need to deal with.
From a user's point of view MCS is very similar to the usage of a
classic Unix shell, that is a command line interface with a prompt on
which users can execute commands in their own environment and wait for
the output before a new command is issued. It is therefore possible to
make a comparison between the ``components'' of a shell, and those
of an MCS connection:
\begin{center}
\begin{tabbing}
      programs, shell scripts        \= $\rightarrow$ \= external programs (\verb|EXEC| command) \kill
      \textbf{Unix shell} \>               \> \textbf{MCS server} \\

      \verb|stdin| and \verb|stdout| \> $\rightarrow$ \> bidirectional TCP socket \\
      system account                 \> $\rightarrow$ \> MySQL account \\
      internal commands              \> $\rightarrow$ \> base commands \\ 
      programs, shell scripts        \> $\rightarrow$ \> external programs (\verb|EXEC| command) \\ 
      home directory                 \> $\rightarrow$ \> work directory
\end{tabbing}
\end{center}
Sent/received data are binary formatted using the MCS protocol and then
the output of a command won't be ASCII text like in a shell.
%

Deriving a C++ class means creating a new class that preserves all the
characteristics of the parent class but with
more specific behaviors added. In the MCS case, the server
behavior can be customized through the derivation of the
\verb|UserThread| class (as described in the documentation, only the
``virtual'' methods should be overloaded).
This way it is possible to create
custom commands available as if they were ``base commands''.
Another class that can be derived for
customization is the \verb|LocalThread|. It runs in a server side thread,
independently from other client threads. This class can be used to
implement some server side tasks, like data quick look or reduction,
DB maintenance, etc.

\section{Other Uses of the MCS Library and the Companion Tools}
The MCS library contains several classes to deal with different
tasks. These classes are mainly used by the MCS server itself, but can
also be used without a running MCS server. Some of the
tasks covered by these classes are:
 threads and synchronization,
 sockets,
 parsing of command lines,
 DB access,
 VOTable and FITS read/write.

It is possible to execute queries on a (MySQL) DB server
using the \mbox{\textbf{DBConn}}, \textbf{Query}, \textbf{Record}
and \textbf{Table} classes. All these classes can be used not only
with C++ programs, but also with all those programming languages for
which we have an interface: C, Fortran, IDL, PHP, Python (Java, Perl
and Tcl/Tk will be available soon).

The \textbf{VOT\_Parser\_Table} class provides an easy interface to read
VOTable and FITS files at the highest level.
At lower level \textbf{VOT\_Parser\_Stream} can be used to read a stream
(like a SAX parser) and \textbf{VOT\_Parser\_Tree} to read a tree in memory
(like a DOM parser). Of course, also these classes
are available to languages for which we have an interface.
\emph{For all the details please see the documentation on the web site!}

MCS is accompanied by a number of software tools to be used
with the MySQL DB server to improve its functionalities. These
tools, once installed, become part of the DB server itself and
the facilities they provide are immediately and transparently available to
anyone who connects to that server with no need to write
a single line of code: \\
\textbf{MyRO} (My Record Oriented privilege system): it
  provides a natural extension to the MySQL privilege system,
  offering the possibility to specify privileges (user/group) on a record
  level (requires MySQL $\ge 5.0$); \\
\textbf{DBEngine for FITS and VOTable}: it
  gives MySQL the ability to read/write DB tables in VOTable and FITS format,
  which means that you can perform queries on them, as well as
  write into them as a result of a computation from other tables (under
  development); \\
\textbf{Database indexing using HEALPix and HTM}:
  this software provides indexing facility for data with spherical
  coordinates allowing, e.g., very fast circular and
  rectangular selections of entries in very large tables (under development).

\section{Conclusions and the Future}
We believe MCS is an extremely useful tool and it is in continuous development.
It will soon implement interfaces for other languages such as
Perl, Java, Tcl/Tk. Another
feature that will be implemented soon is the integration with MyRO to
handle a more complex privilege system.
User contributed libraries for the various supported languages are
being built. This will allow an even easier access to the MCS
functionalities to non expert programmers. Moreover, commonly used,
independently developed external packages will also be included and
made accessible trough MCS. They include the afore mentioned HEALPix
and HTM libraries for sky pixelization scheme, the Naval Observatory
Vector Astrometry Subroutines (\htmladdnormallink{NOVAS}
{aa.usno.navy.mil/software/novas}) used for astrometric calculations and
transformations, the World Coordinate System (\htmladdnormallink{WCS}
{fits.gsfc.nasa.gov/fits_wcs.html}) library and tools.

\end{document}